# Magnetization Effects in Proton Micro-Irradiated Diamond

T. N. Makgato[1, a], E. Sideras-Haddad[1, 4], S. Shrivastava[1], M. Madhuku[2], K. Sekonya[1, 2], C. A. Pineda-Vargas[3,5], D. Joubert[1, 4]

1. School of Physics, University of the Witwatersrand, Johannesburg 2050, South Africa

2. iThemba Laboratory for Accelerator Based Sciences, Gauteng, Johannesburg 2050, South Africa

3. Materials Research Department, iThemba LABS, P. O. Box 722, Somerset West 7129, South Africa

4. Center of Excellence in Strong Materials, University of the Witwatersrand, Johannesburg 2050, South Africa

5. Faculty of Health and Wellness Sciences, Cape Peninsula University of Technology, P. O. Box 1906, Bellville 7535, South Africa

**ABSTRACT.**

We report the magnetic response of high purity type IIa single crystal CVD diamond following high energy irradiation by 2.2 MeV proton micro-beams, produced using Nuclear Microprobe techniques. Post-irradiation microanalysis is achieved using Atomic Force, Magnetic Force and Electrostatic Force Microscopy (AFM, MFM & EFM) as well as Raman spectroscopy. Comparison between MFM and EFM measurements at zero electrical bias is used to preclude artefacts of electrostatic origin suggesting the presence of magnetic states in the microscale regions induced by proton micro-irradiation. Analysis of the irradiated regions using probe polarization dependent magnetization measurements shows an irradiation induced magnetic response consistent with ferromagnetic ordering at room temperature. Structural characterization of the micro-irradiated regions using $\mu$-Raman spectroscopy indicates a correlation between formation of the $sp^2$ carbon phase in the diamond matrix with the scale of the observed magnetization, in the investigated experimental regime.

Recent experimental reports of room temperature magnetic ordering in organic materials have sparked considerable interest in studies of magnetic systems in carbon based materials.

---

[a] **Corresponding Author**
Email: Thuto.Makgato@students.wits.ac.za

Although the phenomenon of room temperature ferromagnetism in organic materials with Curie points $T_C$ above room temperature was postulated much earlier following studies by Turek et al.[1] and Allemand et al.[2], thorough systematic studies of impurity contributions and reproducibility of magnetic signatures were necessary to support the existence of such phenomena. X-ray circular magnetic dichroism (XMCD) studies have been used to show that magnetic ordering in proton micro-irradiated HOPG films is primarily correlated to π-electronic systems and is sensitive to various experimental parameters such as the ion fluence and charge density[3]. The weakness of magnetic signatures characteristic of irradiated carbon systems has often created much scepticism in the scientific community. Sepioni et al.[4] showed that agglomeration of ferromagnetic impurities in typical HOPG samples can trigger ferromagnetic ordering of several $\mu$emu in magnitude, comparable to signals observed from various irradiated samples[5-7]. Nonetheless, since background measurements using SQUID prior irradiation are routinely carried out and subtracted from results, in addition to other quantitative techniques such as PIXE and RBS it is unlikely that these observations can be attributed to magnetic impurities, instead, it is the effects of irradiation that are quantified with <$\mu$emu sensitivity[7].

The origin of irradiation induced magnetic ordering is still a subject of extensive investigation. Various structures attributed to the origin carbon magnetism include; carbon interstitials[8], zigzag edges[9-10], negative Gaussian curvature[11] and vacancies[12-15], nonetheless, no general consensus as has been adopted in the scientific community. Using spin polarized DFT studies Zhang et al.[15] showed that when a single vacancy is introduced into the diamond lattice, the structure undergoes a Jahn-Teller distortion resulting in a stable magnetic moment of 2 $\mu_B$ at room temperature, compared to 1.4 $\mu_B$ in the case of graphene[14]. The existence of various impurities such as N in close proximity to a single vacancy in diamond can result in a higher magnetic moment at room temperature. In principle, it is thus easier to induce magnetic ordering in diamond than in HOPG or graphene, however, apart from irradiation experiments of nanodiamonds using nitrogen beams[16], few reports of irradiation induced magnetism studies in diamond systems are found in literature. Carbon materials are particularly interesting in that they can be incorporated into organic systems such as biological tissue for biomedical applications. Numerous studies have shown that carbon systems such as graphene, carbon nanotubes, and $C_{60}$, exhibit outstanding physical and chemical properties that could provide critical solutions to several problems in science from both a fundamental point of view and for a variety of applications[17-20]. Extreme properties of diamond including superior high thermal conductivity, mechanical ultra-hardness, radiation hardness, high electron-hole mobility and wide transparency to electromagnetic probes from ultraviolet to infrared frequencies, render diamond based materials viable for a vast range of applications[21].

Recent studies have shown that the Nitrogen-Vacancy centre (NV) has a weak spin-orbit coupling in the diamond matrix[22,23]. This results in long spin coherence times at room temperature, long diffusion lengths and allows for fast resonant spin manipulation. These properties among several others, demonstrate the suitability of diamond as a host material for fabricating scalable qubits as building blocks for quantum information devices[24-26]. In addition, Dutt *et al.*[27] demonstrated that the $^{13}C$ nucleus can be used as a quantum register by coupling to individual electron spins associated with the NV colour centres which can be coherently manipulated using optical and microwave frequencies. Since pure insulating diamond can be efficiently doped into a semiconducting state, the ability to induce magnetic ordering in diamond could present the opportunity to create a controllable localized magnetic environment for electron and nuclear spins associated with the NV colour centres to achieve practical spintronics and advanced quantum information devices. In the present study type IIa ultra-pure single crystal detector grade diamond samples were investigated as potential candidates for radiation induced magnetic ordering.

Proton beams were accelerated by the 6MV Tandem accelerator at iThemba LABS (Gauteng, South Africa) to an energy of 2.2 MeV and were then channelled by a series of electrostatic elements into the Nuclear Microprobe apparatus where proton micro-beams of ~1-10 $\mu m$ spatial resolution are produced. Similar experiments were conducted at iThemba LABS (Somerset West, South Africa) using an Oxford triplet Nuclear Microprobe system linked to a 6 MV single ended Van der Graff accelerator[28]. The diamond samples were then irradiated under normal incidence onto clean (100) surfaces using the produced 2.2 MeV proton micro-beams. As a test for reproducibility, two sets of experiments were carried out, the first on sample 1 using the Tandem based Nuclear Microprobe and the second on sample 2 (FIG. 1) using the Oxford Triplet Nuclear Microprobe.

Particle Induced X-Ray Emission (PIXE) as well as Rutherford Backscattering Spectroscopy (RBS) techniques incorporated into the Nuclear Microprobe chambers were utilized to assess the fraction of intrinsic magnetic impurities in the diamond samples introduced during chemical vapour deposition (CVD) synthesis at DeBeers Element Six. The total charge deposited in the micro-irradiated regions ranged between 20 nC and 510 $\mu$C while fluences from $2\times10^{16}$ $H^+/cm^2$ up to $1.4\times10^{18}$ $H^+/cm^2$ were explored as tabulated in TABLE I. Surface areas ranging from 650 $\mu m^2$ up to $3.1\times10^4$ $\mu m^2$ were irradiated. SRIM (SRIM2013) Monte Carlo Simulations[29] for protons of 2.2 MeV in diamond estimate a longitudinal range of ≈28 $\mu$m with a longitudinal straggle of ≈0.25 $\mu$m where the diamond displacement energy is set to 45 eV to account for the radiation hardness nature of diamond following studies of displacement energy in diamond[30,31].

Following proton irradiation of the diamond (100) surfaces, a Veeco dimension 3100 AFM, driven with a Nanoscope IIIa controller operating in tapping/lift dual mode was used to characterize the morphology as well as local magnetic and electrostatic force field gradients of the irradiated micro-regions. Several probes with metallic alloy coatings were utilized in order to investigate the variation of the observed magnetic and electrostatic force field gradients across the samples. The MFM probes utilized for AFM/MFM characterization are commercially available MESP-type probes of high magnetic moment with ≈10-150 nm Co/Cr ferromagnetic alloy coating on the front and back sides of the cantilever. To investigate the magnetic response and preferential alignment of the magnetic moments in the irradiated microscale regions, we compared data acquired using probes magnetized in opposite directions (i.e. into the diamond (100) plane and out of the diamond (100) plane) as well as non-magnetized probes (as synthesised). The MFM probes were magnetized over similar time intervals under the same magnetic field. In general, lateral resolution in MFM imaging using standard probes is approximately equal to the lift scan height[32]. However, recent studies geared towards improving the lateral resolution of MFM probes for research and specialized industrial applications resulted in the fabrication of MFM probes with ≈10-30 nm lateral resolution[33-35].

In addition to the magnetic MFM probes, we utilized commercially available non-magnetic probes of type SCM-PIT with ≈20 nm PtIr coatings on the front and back sides of the cantilever in order to assess possible interference to the MFM signals. Measurements acquired under different electrical bias conditions were then used to compare acquired EFM data with MFM data. To investigate the structural properties of pristine and micro-irradiated diamond regions, we carried out $\mu$-Raman spectroscopy measurements using the 514 nm emission line from an argon laser in conjunction with a Horiba Jobin-Yvon LabRaman HR spectrometer. The incident laser beam was focused onto the samples using the Olympus microscope attachment with a 50X objective lens. The backscattered light was dispersed via a 1800 lines/mm grating onto a liquid nitrogen cooled charge coupled device (CCD) detector.

In-situ PIXE and RBS results show that the diamond samples contained magnetic impurities below the 10 ppm level. The detected intrinsic magnetic impurities included Cr, Fe and Ni. Optical images of diamond (100) surfaces following proton irradiation are shown in FIG. 1, with visible radiation damage over the different irradiated micro-regions and for all beam fluences used. FIG. 2 shows an example of a TM-AFM topography image of region-A of Sample-1 together with the corresponding line profile. For all investigated regions, the radiation induced surface up-swelling observed using TM-AFM is consistent with designated surface area of the micro-irradiations and is generally proportional to the ion fluence. Surface roughness measurements (rms.) for both samples yielded $R_q \approx$1-5 nm as deduced from TM-AFM analysis.

Maximum radiation induced damage in the diamond occurs near the end of range (EOR) of the protons ($\approx$28 $\mu$m) where the nuclear stopping force dominates the electronic stopping force[36] producing on average $\approx$11 vacancies/ion according to SRIM2013 calculations. A relatively undamaged confining cap therefore exists along the ion path between the surface and EOR damaged region. This results in the formation of a microscale region under high pressure with low probability of diamond-to-graphite structural relaxation in the bulk[37].

During MFM imaging, the cantilever is driven into oscillation at or near its resonance frequency. Interaction of the oscillating MFM probe with the magnetic field generated from the sample under investigation results in a measurable phase shift in the oscillating cantilever. The dependence of the measured phase shift on the force gradient can be expressed as:

$$\Delta\Phi \approx -\frac{Q}{k}\frac{\partial \boldsymbol{F}}{\partial z} = -\mu_0 \frac{Q}{k}\left[-q\frac{\partial H_z}{\partial z} + m_x \frac{\partial^2 H_x}{\partial z^2} + m_y \frac{\partial^2 H_y}{\partial z^2} + m_z \frac{\partial^2 H_z}{\partial z^2}\right] \quad (1)$$

where $m_i$ and $H_i$ with $i = x,y,z$ are the cartesian components of the tip dipole moment **m** and of the stray-field **H** of the sample, $\mu_o$ is the magnetic permeability of free space, Q is the quality factor, $k$ is the spring constant of the oscillating cantilever and ***F*** is the magnetic force vector[38]. Without accurate knowledge of the probe geometry and magnetic properties, Eq. (1) is only useful in conducting qualitative assessment of the sample's magnetic response. FIG. 3 shows an example of a MFM Phase shift image of region-A of Sample-1 acquired from a lift scan height of 50 nm as well as a line profile extracted across the region indicated in the phase image. The magnetic probe magnetized in direction pointing into the (100) diamond plane ('+1') detected phase shift signals up to ≈1 deg. Phase shift signals recorded from region-A of sample-1 shown in FIG. 3 are the highest of all the investigated micro-irradiations and represent the micro-scale region with the highest fluence utilized in the present study, which corresponds to ≈$1.4 \times 10^{18}$ $H^+/cm^2$ (see TABLE I). Assuming an experimental error of ≈ 0.1 deg. in the phase shift signal and given that the cantilever spring constant $k$ of the magnetic probe is ≈ 4.3 $Nm^{-1}$ and the quality factor Q ≈ 310, the magnetic field gradient $\partial F/\partial z$ measured in region-A of Sample-1 shown in FIG. 3 thus ranges between (1.4 and 14) pN/nm in accordance with Eq. (1).

MFM investigations of all micro-scale irradiated regions show that the ion fluence plays an important role in the observed magnetic response. Fluences below 8.4 $\times 10^{17}$ $H^+/cm^2$ displayed a very weak magnetic response in the phase signal. This suggests that the observed magnetic response may be closely related to the spacing between point defects such as single vacancies created in the irradiated region. During scanning, the MFM probe interacts with a range of forces generated from the sample. From approximately 50 nm and higher above the surface, long range forces prevail while short range forces become negligible. FIG. 4 shows the magnetic response of region-A of Sample-1 as function of the lift scan height (nm) with the magnetic probe magnetized in the '+1' direction. The strength of the magnetic field is inversely proportional to the lift scan height and decreases rapidly with increasing lift scan height. FIG. 4 shows that the magnetic signal measured over the micro-irradiated region is consistent and stable even up to 200 nm above the surface. As region-A Sample-1 displays the most intense phase shift contrast in all the micro-irradiations, this region was further investigated using non-magnetized probes '0', as well as probes magnetized in the '+1' and '-1' magnetization directions.

FIG. 5 shows phase shift (deg.) profiles of region-A of Sample-1 acquired at a lift scan height of 50 nm using different magnetizations, namely '+1', '-1' and non-magnetized '0' (i.e. as fabricated). As shown in FIG. 3 and 4, negative phase shifts were observed in the micro-irradiated region for all the probe magnetizations (FIG. 5). The micro-irradiated regions appear to generate a measurable magnetic field in which the orientation of the net magnetic moments is aligned in the same direction irrespective of the magnetic field direction applied to the magnetic probe. The strength of the observed magnetic response is dependent on the magnetic field orientation with the strongest and weakest phase shift signals observed in the '+1' and '0' magnetization directions respectively. This indicates that the application of a magnetic field to the MFM probe influences the orientation of the magnetic moments in the micro-irradiated regions. However, both electrostatic and magnetic fields are long range forces detected by virtue of phase shifts from the piezo drive, therefore local electrostatic field gradients can in principle couple with magnetic field gradients in empirical observations resulting in convoluted information and the possibility of measuring artefacts.

Probes utilized in the present MFM study are both ferromagnetic and electrically conductive. Since electrostatic fields are in principle measurable above 50 nm, it is necessary to investigate the behaviour of electrostatic forces in the micro-irradiated regions before any reasonable conclusions can be deduced from such empirical analysis. In the present study,

EFM images were acquired initially using Co/Cr coated MESP-HM probes and then using PtIr coated type SCM-PIT probes under electrical bias voltages ranging between 0V and 10V. Again for brevity we present only the data acquired from region-A of Sample-1 where the highest proton fluence was used. FIG. 6 shows an example of an EFM image acquired over irradiated region-A of Sample-1 using a commercially available SCM-PIT type probe subjected to an electrical bias of 4V at a lift scan height of 50 nm. The EFM image shows negative phase shifts associated with attractive forces in the micro-irradiated region as expected from Eq. (1) as in the observed MFM images (FIG. 3). According to *Cherniavskaya et al.*[39], the electrostatic force acting between the probe and the sample can be written as:

$$F(z) = \frac{1}{2}\frac{dC_{s-t}}{dz}V_t^2 + E_S Q_t \qquad (2a)$$

where $C_{s-t}$, $V_t$, $E_S$ and $Q_t$ refer to the tip-sample capacitance, the voltage applied to the tip, the electric field at the tip location that is only created by the charges and/or multipoles in the sample surface and the effective charge on the tip (where; $Q_t = C_{s-t}V_t + Q_{im}$ where $Q_{im}$ is the image charge in the tip induced by the static charge distribution on the sample surface) respectively. The voltage dependence of the electrostatic field gradient measured using type MESP-HM and SCM-PIT probes at a constant lift scan height of 50 nm is presented in FIG. 7 by virtue of the cantilever phase shift.

Although all probes used in this investigation are sensitive to electrostatic forces and are capable of detecting very low electrostatic signals and in special cases even single charge events[40], only type MESP-HM Co/Cr coated probes are sensitive to both magnetic fields as well as electrostatic fields that exist between the probe and the samples. EFM data acquired using both probe types confirm the expected quadratic voltage dependence of the electrostatic force gradient (FIG. 7) on the applied electrical voltage at least up to around 6V above which the EFM response appears to deviate from the normal quadratic behaviour (see FIG. 7 insert).

Application of quadratic polynomial functions to the acquired EFM data (FIG. 7) yields $R^2$ and adjusted $R^2$ values of ≈0.99 for all probe types.

It is expected that the electric field at the tip $E_S$ as well as the image charge $Q_{im}$ are approximately constant in the surface normal direction given the dimensions of the probe tip relative to the micro-irradiated area, the low surface roughness ($R_q$≈ 1-5 nm) as well as the constant lift scan height technique applied during imaging. Apart from verifying the functional quadratic dependence of the electrostatic field gradient on the applied electrical bias (Eq. (2a)), the quadratic polynomial function fitted to data acquired using SCM-PIT probes also show that the electrostatic contribution at zero electrical bias is ≈0V, i.e. $\partial(E_S Q_{im})/\partial z \approx 0$ as expected since both $E_S$ and $Q_{im}$ are approximately constant in the scan configuration as discussed above. Since the cantilever spring constant $k$ is known and the probe quality factor $Q$ can be readily calculated during cantilever tuning, using Eqs. (2a) and (1) as well model free parameters we can estimate mathematical terms of the electrostatic force gradient where we have the electrostatic contribution during EFM measurements given by:

$$\Delta\phi \approx -\frac{Q}{k}\left[\frac{\partial^2 C_{S-t}}{\partial z^2}\frac{V_t^2}{2} + \frac{\partial}{\partial z}E_S C_{S-t}V_t + \frac{\partial}{\partial z}E_S Q_{im}\right] \quad (2b)$$

Therefore, EFM data acquired using PtIr coated SCM-PIT probes reveal no significant phase shift at zero electrical bias. Since SCM-PIT probes are not sensitive to magnetic fields at room temperature, this observation suggests that any electrostatic fields resulting from possible variations in $E_S Q_{im}$ at zero electrical bias are either not detectable or are negligible using the current experimental setup. On the contrary, voltage dependent EFM measurements acquired using MESP-HM probes reveal phase shifts up to ≈1 deg. at zero electrical bias (FIG. 7).

The observed scales are closely comparable with MFM data acquired using MESP-HM probes under no electrical bias (FIG. 5). Since EFM data acquired using SCM-PIT probes at zero electrical bias demonstrate that the image charge as well as tip electric field induced by the charges and/or multipoles in the diamond sample surface are not sufficient to generate measurable electrostatic field gradients at a lift scan height of 50 nm (and above 50 nm given that field is inversely proportional to the lift scan height), it is therefore reasonable to deduce that the phase shifts observed using MESP-HM probes do not originate from electrostatic fields but rather from local magnetic fields in the proton micro-irradiated regions. The scale of the phase shift signal is primarily dependent on the magnitude of magnetization and the magnetization direction of the MFM probe as shown in FIG. 5 and is also comparable with phase shift magnitudes observed in previous studies of proton irradiation in other carbon materials[5,6,41]. In FIG. 5 the MFM response of the proton micro-irradiated region-A of Sample-1 shows that the magnitude of the observed magnetic field gradient increases significantly following probe polarization in the directions normal and anti-normal to the (100) diamond surface. This behaviour has been described to be an indication of a possible ferromagnetic state in the irradiated region[5,6,41]. It has been shown that magnetic ordering in carbon materials is strongly related to spin polarization in the π-electron systems which are typically associated with the $sp^2$ phase of carbon. In FIG. 8 µ-Raman spectroscopy measurements acquired over sample 1 reveal structural properties of pristine, and several micro-irradiated regions in diamond. In FIG 8, the Raman signature of pure diamond showing the diamond peak (1331 cm$^{-1}$) is used to assess the diamond quality (FWHM) and also serves as reference to assess the structural changes induced by micro-irradiation. For brevity, we compare the effects of micro-irradiations A, C and E (see Table I and FIG 1) from sample 1. In micro-irradiations C and E, radiation damage in the diamond structure results in significant reduction of the diamond quality as noted from the broadening and reduction in the intensity of the diamond peak. In addition, micro-irradiations C and E show a downshift in the diamond line which indicates irradiations induced mechanical stress as well asymmetric broadening in the case of C, which could be indicative of quantum mechanical effects such as phonon confinement. However, the structure in regions C and E remain primarily in the $sp^3$ phase. The Raman signature for micro-irradiation A is particularly different. Experimental parameters (beam confinement) selected for micro-irradiation A allow the formation of π-electronic states and transformation of $sp^3$ phases into $sp^2$ phases of carbon upon irradiation. Upon visible excitation (514 nm), π-electronic states created in the micro-irradiated region A during phase transformation resonates with visible excitation hence the Raman signature is dominated by $sp^2$ sites. This is clearly evidenced by the emergence of a distinct D-peak (1360 cm$^{-1}$) which is due to the breathing modes of $sp^2$ atoms in rings[42-44]. In addition, no significant broadening or shift in the diamond peak is observed. Therefore, the energy imparted into the micro domain A upon proton irradiation is expended primarily in phase transformation processes and only partial amorphization. The micro-domain thus consists of both $sp^3$ and $sp^2$ phases of carbon.

Although a critical ratio of $sp^2$ to $sp^3$ chemical bonding has been attributed to the emergence of carbon ferromagnetism in other studies[45], theoretical investigations based on DFT methods have also indicated the possibility of inducing ferromagnetism in carbon by point defects, in particular, vacancies[14,15] where the vacancy concentration (hence vacancy spacing) can be used as a control parameter. The fluences explored in the present study (TABLE I) imply (see Ramos *et al.*[7]) that the vacancy spacing for all micro-irradiations falls within the narrow range [0.3; 4.5] nm. However, their characteristic Raman signatures as well as magnetic signals measured by MFM reveal appreciable differences in detail. Therefore experimental parameters other than the vacancy spacing e.g. charge density, may play an important role in modifying the bonding environment (e.g. by phase transformations) within which the point defect (vacancies) are embedded to trigger long range magnetic order. In addition, irradiation carried using a micro beam results in pronounced geometrical modification of the local irradiated surface (up to ~100-150 nm upswelling is observed) which can in principle trigger magnetic ordering by negative Gaussian curvature[11]. Since in principle, vacancy induced magnetic order can exists in diamond ($sp^3$) and $sp^2$ carbon, it is possible that the overall magnetic signals may be of a complex nature i.e. due to a combination of multiple structures (e.g. vacancies in both $sp^2$ and $sp^3$ carbon). However, providing empirical estimates of the experimental parameters necessary to trigger ferromagnetism is therefore not trivial and would require a reasonably large sample space which could be experimentally challenging. Nonetheless, in the present study we have demonstrated using a limited set of experimental parameters, the formation and observation of magnetic domains with ferromagnetic-like behaviour at room temperature in ultra-pure type IIa CVD diamond following 2.2 MeV proton micro-irradiation. Given the numerous possible applications, investigation into the origin and optimization of the observed magnetic response in diamond is imperative and requires further extensive studies.

# ACKNOWLEDGEMENTS

This research is funded by the Center of Excellence in Strong Materials and the University of the Witwatersrand in South Africa. We would also like to acknowledge the contribution by Dr Rudolph Erasmus of the University of the Witwatersrand for his assistance with the acquisition of Raman spectroscopy measurements.

# TABLES

**TABLE I.** Irradiation summary of IIa-diamond (100) surfaces by 2.2 MeV Proton micro-beams.

| | Sample 1 | | | Sample 2 | |
|---|---|---|---|---|---|
| Region | Fluence (x $10^{17}$ $H^+$/cm$^2$) | Charge ($\mu$C) | Region | Fluence (x $10^{17}$ $H^+$/cm$^2$) | Charge (nC) |
| A | 14 | 8.1 | A | 0.2 | 21 |
| B | 10 | 7.5 | B | 0.74 | 78 |
| C | 1.1 | 2.0 | C | 2.1 | 200 |
| D | 2.1 | 4.0 | D | 8.4 | 900 |
| E | 10 | 50 | E | 3.8 | 2200 |

# FIGURES

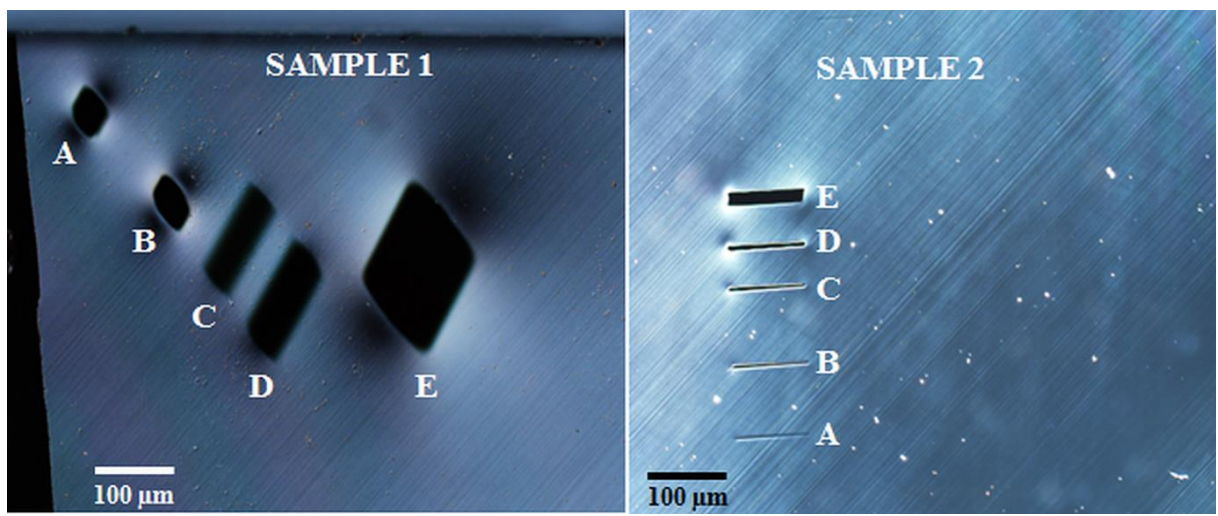

**FIG. 1**. Optical images of micro-irradiations on type IIa diamond (100) surfaces by 2.2 MeV Proton micro-beams.

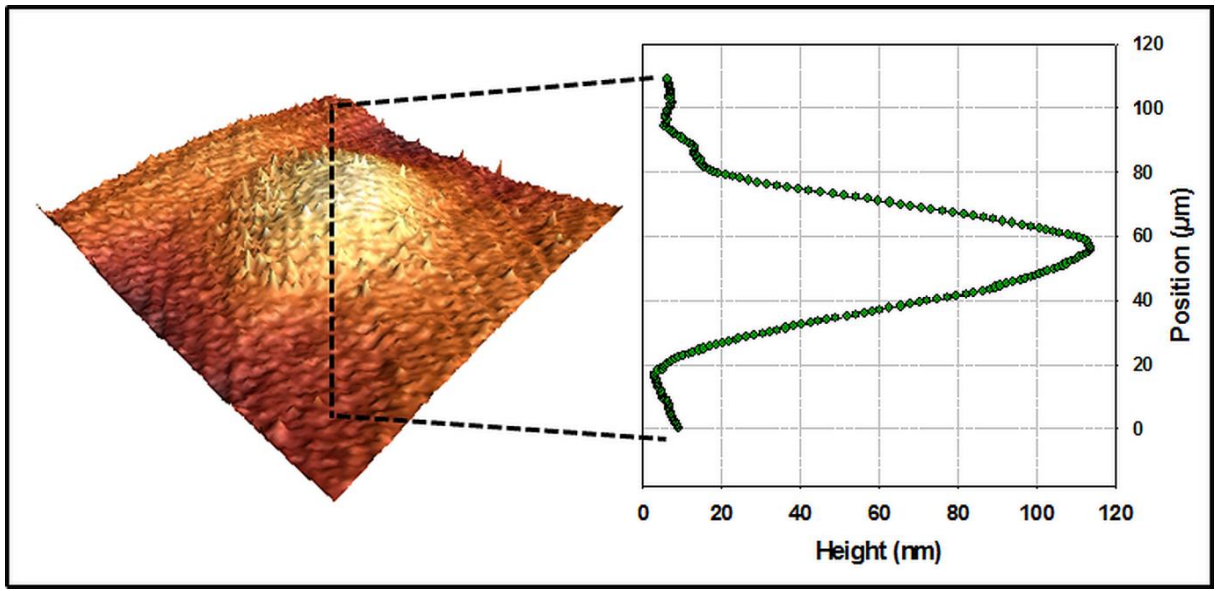

**FIG. 2**. Example of a Tapping Mode AFM topography image acquired over region-A of Sample-1 together with a line profile extracted across the region indicated by the dotted and solid lines.

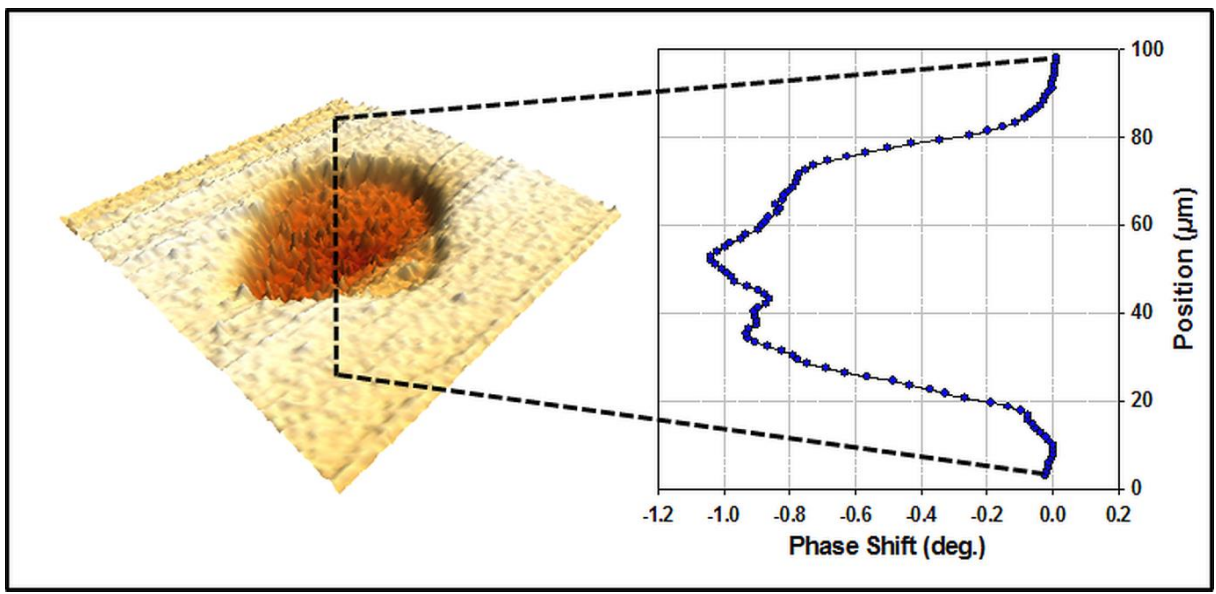

**FIG. 3**. Example of a MFM Phase image (deg. x10$^{-3}$) of region-A sample-1 acquired 50 nm above the diamond (100) surface together with the corresponding line profile (deg.) extracted from the region indicated by the solid and dotted lines.

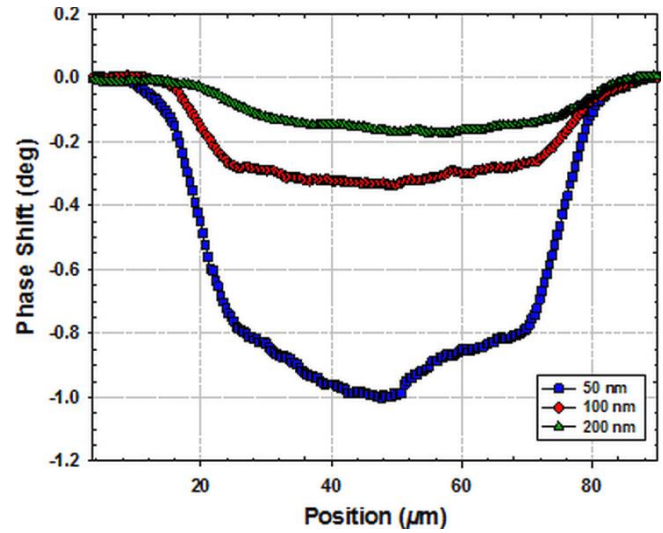

**FIG. 4**. (Colour online) Phase shift (deg.) versus Position (μm) of the magnetic probe over region-A sample-1 at several lift scan heights (nm) with the probe polarized in the '+1' polarization direction.

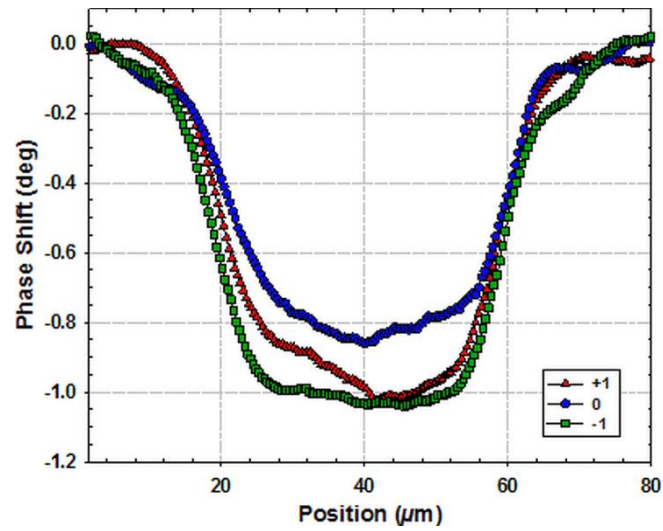

**FIG. 5.** (Colour online) Phase shift (deg.) versus Position (μm) of the magnetic probe over region-A of sample-1 under different probe magnetization directions.

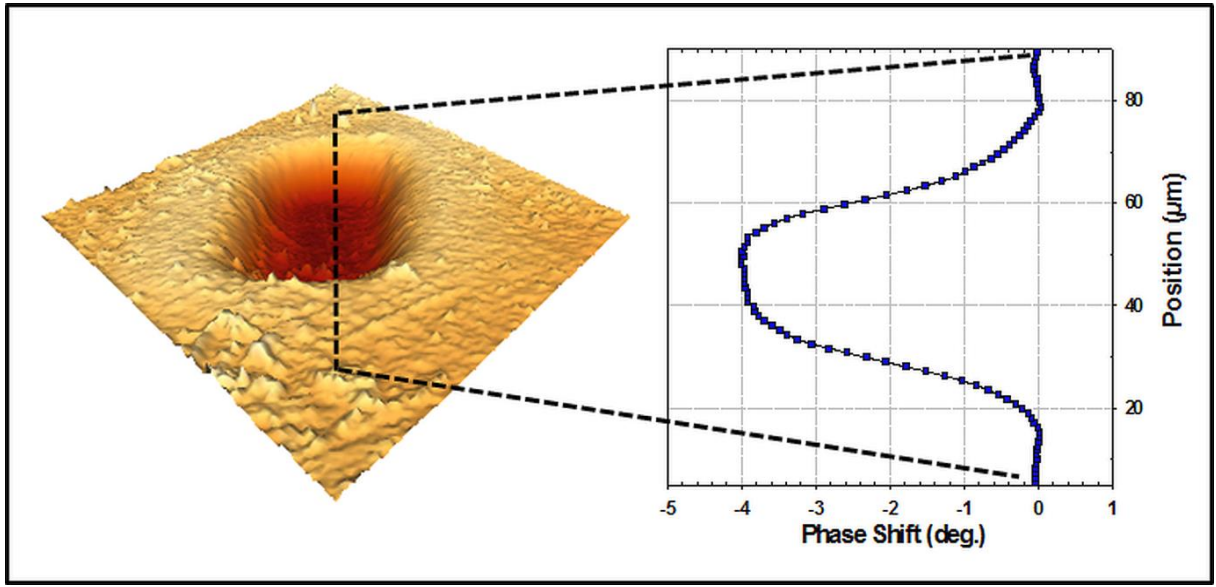

**FIG. 6**. Example of an EFM image acquired over region-A of sample-1 using a SCM-PIT type probe electrically biased at 4V together with the corresponding line profile (deg.) extracted across the region indicated by the dotted lines.

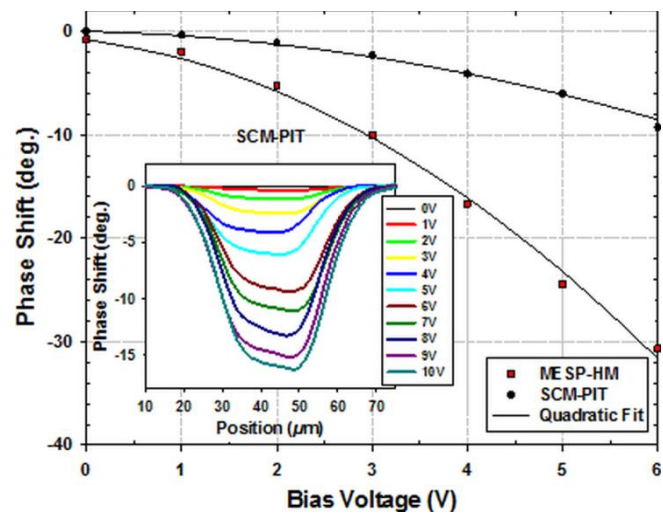

**FIG. 7.** (Colour online) Electrostatic response measured as a phase shift (deg.) from the piezo drive at 50 nm above region-A of Sample 1 using MESP-HM (■) and SCM-PIT (●) probes types as a function of the applied electrical bias voltage (V). The insert shows an example of EFM profiles acquired using SCM-PIT probes illustrating the deviation from normal quadratic behaviour at higher bias voltages.

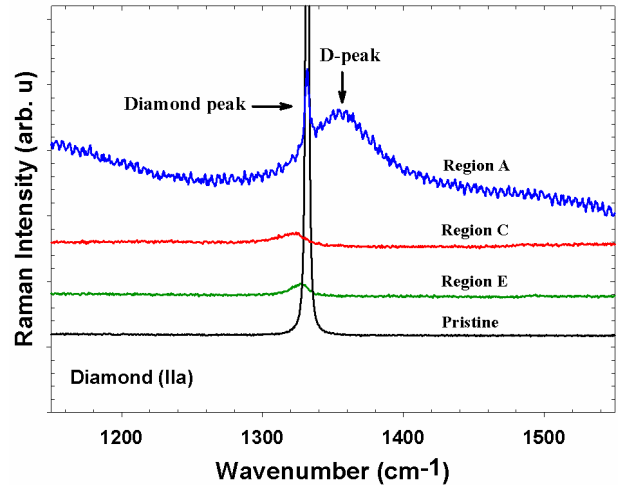

FIG. 8. *μ*-Raman spectroscopy of diamond following micro-irradiation using 2.2 MeV protons. The Raman spectra are acquired from the pristine and micro-irradiated regions A, C and E of sample 1 for comparison.